\documentclass[a4paper,12pt]{article}
\usepackage{setspace}

\usepackage{amsfonts,amscd,amssymb,amsmath,amsthm,latexsym}
\usepackage{tabularx}
\newcolumntype{C}{>{\centering\arraybackslash}X}

\usepackage{float}
\usepackage{graphicx}
\usepackage{epsfig}

\newcommand{\FF}{\mathcal{F}}

\def\bs{\begin{small}}
\def\es{\end{small}}
\def\bfs{\begin{footnotesize}}
\def\efs{\end{footnotesize}}
\def\bi{\begin{itemize}}
\def\ei{\end{itemize}}

\begin{document}

\centerline{\bf \Large The Schwarzschild-de Sitter Metric}

\medskip

 \centerline{\bf \Large of Nonlocal $\sqrt{dS}$ Gravity}

\bigskip
\bigskip

\centerline{\bf Ivan Dimitrijevic$^1$, Branko Dragovich$^{2,3,4}$, Zoran Rakic $^1$ and Jelena Stankovic$^5$}

\bigskip

$^{1}$Faculty of Mathematics,  University of Belgrade,  Studentski trg 16,  11158 Belgrade, Serbia \\
$^{2}$Institute of Physics,  University of Belgrade, 11080  Belgrade, Serbia \\
$^{3}$Mathematical Institute of the Serbian Academy of Sciences and Arts, 11000 Belgrade, Serbia \\
$^{4}$Laboratory for Theoretical Cosmology, International Centre of Gravity and Cosmos, Tomsk State University of Control Systems and Radioelectronics (TUSUR), 634050 Tomsk, Russia \\
$^{5}$Faculty of Education,  University of Belgrade,  Kraljice Natalije 43, 11000  Belgrade, Serbia

\bigskip

\centerline{Abstract}

\medskip

{It is already known that a simple nonlocal de Sitter gravity model, which we denote as  $\sqrt{dS}$ gravity, contains an exact vacuum  cosmological solution which mimics dark energy and dark matter and is in  very good agreement with the standard model of cosmology. This success of $\sqrt{dS}$ gravity motivated us  to investigate how it works at lower than cosmic scale -- galactic and the solar system.  This paper contains our investigation of the corresponding Schwarzschild-de Sitter metric of the $\sqrt{dS}$ gravity model. To get exact solution, it is necessary to solve the corresponding nonlinear differential equation, what is a very complicated and difficult problem. What we obtained is a solution of linearized equation, which is related to space metric  far from the massive body, where gravitational field is weak.  The obtained approximate solution is of particular interest for examining the possible role of non-local de Sitter gravity $\sqrt{dS}$ in describing the effects in galactic dynamics that are usually attributed to dark matter. The solution has been tested on the Milky Way and the spiral galaxy M33 and is in good agreement with observational measurements. }



\section{Introduction}
The general theory of relativity (GR) \cite{wald} is considered  as one of the most successful and beautiful physical theories. It is worth mentioning the following main predictions and successful confirmations: deflection of light near the Sun,  black holes,  gravitational light redshift, lensing, and gravitational waves, see e.g.  \cite{ellis}.

In the Standard Model of Cosmology (SMC) \cite{robson}, also known as $\Lambda$CDM model, GR is adopted as theory of gravitation at all space-time scales, from the solar system to the galactic and cosmic ones. To describe galactic rotational curves and accelerated expansion of the universe by GR, it was supposed existence of dark matter (DM) and dark energy (DE), respectively. According to the current SMC, the universe matter/energy budget consists approximately of 68 \% of dark energy, 27 \% of dark matter and only 5 \% of standard visible matter. However, despite many experimental and theoretical investigations (as a review, see e.g. \cite{oks}), the existence of DM and DE has not been proven, thus they are still hypothetical constituents of the dark side of the universe.

It should be also mentioned that GR  suffers from the singularities -- the black hole and  Big Bang singularity \cite{novello}. As we know,  if theory  contains a singularity it means its inapplicability when approaching to singularity and that a more general appropriate  theory should be invented. Also, it should be mentioning that GR can not be consistently quantized \cite{stelle}. As we know, other physical theories have their own domain of validity, usually  limited by the  space-time scale and  some parameters, or by complexity of the system.  In this sense, GR should not be an exception and serve as a theory of gravity from the Planck scale to the universe as a whole \cite{clifton}.

Based on all the aforementioned shortcomings, it can be   concluded  that general relativity is not a final theory of gravitation and that there is a sense to look for a more general theory than GR \cite{nojiri}. In principle, there is a huge number of possibilities to extend the Einstein-Hilbert (EH) action and for now there is no rule on how to choose the right path \cite{nojiri1,dimitrijevic1}. Hence, in practice there are many phenomenological approaches and the most elaborated is $f(R)$ theory \cite{faraoni}, where in the EH action the scalar curvature $R$ is replaced by some function $f(R)$. Among  other interesting approaches is the nonlocal one \cite{capozziello,dragovich0}. In the nonlocal gravity models, besides $R$ in the EH action there is a nonlocal term with some invariants usually composed of $R$ and $\Box$, where $\Box$ is the d'Alembert-Beltrami operator.

Depending on how the $\Box$ is built into the nonlocal term, there are mainly two typical  examples of nonlocal gravity models: (i) non-polynomial analytic expansion of $\Box$, see e.g. \cite{biswas2,biswas3,biswas4,biswas5,biswas6,PLB,kolar}, i.e. $\mathcal{F}(\Box) = \sum_{n=1}^\infty f_n \ \Box^n$ (see various examples \cite{koshelev,modesto2,dimitrijevic4,dimitrijevic10,dimitrijevic11,dimitrijevic5}), and (ii) a polynomial of $\Box^{-1}$, see e.g. \cite{capozziello,deser,woodard,maggiore,JHEP}. The motivation for using a local operator of the form (i) is found in string theory -- ordinary and $p$-adic one \cite{dragovich1}. It is obvious that in the case (i) dynamics depends not only on the first and second space-time derivatives but also on the all higher ones. Nonlocal operator $\Box^{-1}$ in (ii)  has its origin in (one-loop)  quantum corrections to some classical field Lagrangians  and is used in investigation of the late time cosmic acceleration without dark energy  \cite{capozziello}.

In several papers, see \cite{PLB,JHEP} and references therein, we investigated the following nonlocal de Sitter gravity model ($\sqrt{dS}$ gravity):
\begin{align} \label{eq1.1}
S = \frac{1}{16 \pi G} \int d^4 x \ \sqrt{-g}\ \big( R-2\Lambda + \sqrt{R - 2\Lambda}\ \mathcal{F} (\Box)\ \sqrt{R - 2\Lambda} \big) ,
\end{align}
where $\Lambda$ is the cosmological constant and nonlocal operator $\mathcal{F} (\Box)$ has the following general form:
\begin{align}  \label{eq1.2}
  \mathcal{F} (\Box) =  \sum_{n=1}^{+\infty} \big(f_n \Box^n  + f_{-n} \Box^{-n}  \big) .
\end{align}
 This nonlocal model is unique compared to other non-local models and its properties will be described in the next section.

It is worth mentioning that \eqref{eq1.1} applied to the homogeneous and isotropic universe gives several exact cosmological solutions \cite{PLB,JHEP}. One of them is $a(t) = A t^{\frac{2}{3}} e^{\frac{\Lambda}{14} t^2} $, which mimics an interplay of dark matter ($t^{\frac{2}{3}}$) and dark energy ($e^{\frac{\Lambda}{14} t^2}$) in  very good agreement with the standard model of cosmology. There are also nonsingular bounce solutions in the flat, closed and open universe as well as singular and cyclic solutions.

The fact that the model works well on a cosmological scale was motivation  to test it on galactic and planetary systems. To this end, it is necessary to get the corresponding Schwarzschild-de Sitter metric. In the paper \cite{Filomat} we presented an initial research with an approximate solution.  In this paper, we provide a much wider and more detailed investigation with some new solutions. It also contains preliminary test  in our galaxy Milky Way and the spiral galaxy M33 with very satisfactory agreement of obtained theoretical results and observational measurements.

This paper is organized as follows.  In Section 2, nonlocal de Sitter gravity $\sqrt{dS}$ is introduced and the equations of motion for the gravitational field are derived. Various aspects of the Schwarzschild-de Sitter metric are presented in Section 3, which in particular contains the solutions and their comparison with observations of the rotation curves of spiral galaxies. Some discussion and presentation of the main results is contained in the last section.

\section{$\sqrt{dS}$ Nonlocal Gravity}

Our nonlocal gravity model is given by its action \eqref{eq1.1}. It can be rewritten in the  compact form

\begin{align} \label{eq2.1}
S = \frac{1}{16 \pi G} \int d^4 x \ \sqrt{-g}\ \sqrt{R - 2\Lambda}\ F(\Box)\ \sqrt{R - 2\Lambda} ,
\end{align}
where $F (\Box)$ is
\begin{align}  \label{eq2.2}
F(\Box) = 1+ \mathcal{F} (\Box) = 1 + \sum_{n=1}^{+\infty} \big(f_n \Box^n  + f_{-n} \Box^{-n}  \big)
\end{align}
with the general form of the d'Alembert-Beltrami operator
\begin{align} \label{eq2.3}
\Box = \nabla_\mu \nabla^\mu = \frac{1}{\sqrt{-g}} \partial_\mu (\sqrt{-g} g^{\mu\nu}\partial_\nu).
\end{align}

 If $F(\Box)= 1$, i.e. $\FF (\Box) = 0$,
then  \eqref{eq2.1} becomes
local de Sitter gravity with action
\begin{align} \label{eq2.4}
  S_0 = \frac{1}{16 \pi G} \int d^4 x \ \sqrt{-g}\ \sqrt{R - 2\Lambda} \ \sqrt{R - 2\Lambda} = \frac{1}{16 \pi G} \int d^4 x \ \sqrt{-g}\ (R - 2 \Lambda) .
\end{align}
 Note that action \eqref{eq2.1} can be easily  obtained from  \eqref{eq2.4}
 by embedding  operator \eqref{eq2.2} inside the product $\sqrt{R - 2 \Lambda}\ \sqrt{R - 2 \Lambda} .$ It should be noted that the degree of $R - 2 \Lambda$ remains unchanged when we go from local \eqref{eq2.4} to non-local action \eqref{eq2.1}, as well as that $F (\Box)$ operator is dimensionless.    It is also worth noting that above local and nonlocal action has the same discrete symmetry, i.e. remains unchanged under transformation $\sqrt{R - 2 \Lambda} \to - \sqrt{R - 2 \Lambda} .$

In this paper, we will not consider the extension of action \eqref{eq1.1} with the matter sector, since we are looking for  the  Schwarzschild-de Sitter metric outside the spherically symmetric massive body.

\subsection{Equations of Motion}

To obtain equations of motion for $\sqrt{dS}$ gravity given by action \eqref{eq1.1}, it is useful to start from more general nonlocal de Sitter model
\begin{align} \label{eq2.5}
S = \frac{1}{16 \pi G} \int d^4 x \ \sqrt{-g}\ \big(R- 2 \Lambda  + P(R)\ \mathcal{F}(\Box)\ Q(R)\big) ,
\end{align}
where  $P(R)$ and $Q(R)$ are some differentiable functions of the Ricci scalar $R .$  Variation of \eqref{eq2.5} with respect to $g^{\mu\nu}$ yields the corresponding  equations of motion (EoM) derived in \cite{variation}, see also \cite{JHEP}.

According to \cite{variation},  the EoM for nonlocal de Sitter gravity model \eqref{eq2.5} are as follows:
 \begin{align} \label{eq2.6}
  G_{\mu\nu} +\Lambda g_{\mu\nu} - \frac 12 g_{\mu\nu} P(R) \mathcal{F}(\Box) Q(R) + R_{\mu\nu}W -K_{\mu\nu}W + \frac 12 \Omega_{\mu\nu} = 0,
 \end{align}
 where $G_{\mu\nu}= R_{\mu\nu} - \frac{1}{2}R g_{\mu\nu} $ is the Einstein tensor, $R_{\mu\nu}$ is the Ricci tensor and
 \begin{align}
 &W =  P'(R)\ \mathcal{F}(\Box)\ Q(R) + Q'(R)\ \mathcal{F}(\Box) P(R) , \quad K_{\mu\nu} = \nabla_\mu \nabla_\nu - g_{\mu\nu}\Box , \label{eq2.7a} \\
  &\Omega_{\mu\nu} =  \sum_{n=1}^{+\infty} f_n \sum_{\ell=0}^{n-1} S_{\mu\nu}(\Box^\ell P, \Box^{n-1-\ell} Q)  -\sum_{n=1}^{+\infty} f_{-n} \sum_{\ell=0}^{n-1} S_{\mu\nu}(\Box^{-(\ell+1)} P, \Box^{-(n-\ell)} Q),  \label{eq2.7b}
  \end{align}
  where
  \begin{align}
  &S_{\mu\nu} (A, B) = g_{\mu\nu} \big(\nabla^\alpha A \ \nabla_\alpha B + A \Box B \big) - 2\nabla_\mu A\ \nabla_\nu B . \label{eq2.7c}
\end{align}
$P'(R)$ and $Q'(R)$ denote the derivative of $P(R)$ and $Q(R)$ with respect to $R$.

Comparing the above equations of motion with respect to their local Einstein counterpart $G_{\mu\nu} +\Lambda g_{\mu\nu} = 0$ they look  very complex and finding some exact solutions  may be a hard problem.

 Since we are interested in the EoM of $\sqrt{dS}$, we have to take $Q(R) = P(R) = \sqrt{R - 2\Lambda}$.  To this end, let us first consider the case $Q(R) = P(R)$. Consequently, equations \eqref{eq2.6}--\eqref{eq2.7c} reduce to
\begin{align} \label{eq2.8a}
  &G_{\mu\nu}+ \Lambda g_{\mu\nu} - \frac{g_{\mu\nu}}{2} P(R) \FF (\Box) P(R) + R_{\mu\nu} W - K_{\mu\nu} W + \frac 12 \Omega_{\mu\nu} = 0 , \\
&W = 2 P'(R)\ \FF (\Box)\ P(R), \quad K_{\mu\nu} = \nabla_\mu \nabla_\nu - g_{\mu\nu}\Box , \label{eq2.8b} \\
&\Omega_{\mu\nu} =  \sum_{n=1}^{+\infty} f_n \sum_{\ell=0}^{n-1} S_{\mu\nu}(\Box^\ell P, \Box^{n-1-\ell} P) -\sum_{n=1}^{+\infty} f_{-n} \sum_{\ell=0}^{n-1} S_{\mu\nu}(\Box^{-(\ell+1)} P, \Box^{-(n-\ell)} P). \label{eq2.8c}
\end{align}

According to our experience, the above equations of motion  \eqref{eq2.8a}--\eqref{eq2.8c} can be significantly simplified and easily solved if there exists a metric tensor $g_{\mu\nu}$ such that for the corresponding d'Alembert-Beltrami operator $\Box$ the following equations (eigenvalue problem) is satisfied:
\begin{align} \label{eq2.9}
\Box P(R) = q\,  P (R), \quad \Box^{-1} P(R) = q^{-1} P(R) , \ \quad  \ \FF (\Box) P(R) = \FF (q)\, P(R) , \qquad q \neq 0,
\end{align}
where $q$ is a parameter of the same dimensionality as  $\Box$.  Applying \eqref{eq2.9} to equations  \eqref{eq2.8a}--\eqref{eq2.8c}, we have
\begin{align}
  &W = 2 \mathcal{F}(q)P'(R) P (R),  \quad \mathcal{F}(q) = \sum_{n=1}^{+\infty} \big( f_n \ q^n + f_{-n} \ q^{-n} \big) , \label{eq2.10a} \\
  &\Omega_{\mu\nu}   = \mathcal{F}' (q) S_{\mu\nu} \big(P , P \big) ,  \qquad \FF'(q) = \sum_{n=1}^{+\infty} n\ f_n \ q^{n-1} -
 \sum_{n=1}^{+\infty} n\ f_{-n} \ q^{-n-1} , \label{eq2.10b}  \\
  &G_{\mu\nu}+ \Lambda g_{\mu\nu} + \FF (q)\Big(  2(  R_{\mu\nu}
  -   K_{\mu\nu} ) P P'   - \frac{g_{\mu\nu}}{2}  P^2(R) \Big)
  + \frac 12 \FF'(q) S_{\mu\nu}(P,P) = 0. \label{eq2.10c}
\end{align}

Now, let us take $P(R) = \sqrt{R - 2\Lambda}$. Then $P'(R) P(R) = \frac{1}{2}$ and $P^2 (R) = R - 2\Lambda.$ Finally, EoM \eqref{eq2.10c} become
\begin{equation} \label{eq2.11}
  \left(G_{\mu\nu}+ \Lambda g_{\mu\nu}\right)\left(1 + \mathcal{F}(q)\right) + \frac 12 \mathcal{F}'(q) S_{\mu\nu}(\sqrt{R-2\Lambda},\sqrt{R-2\Lambda}) = 0.
\end{equation}
If the nonlocal operator satisfies
\begin{align} \label{eq2.12}
\mathcal{F} (q) = -1,  \quad  \quad  \mathcal{F}' (q) = 0 ,
\end{align}
then equations of motion \eqref{eq2.11} are also satisfied.

According to the above consideration, the main problem is to  solve equation $\Box \sqrt{R-2\Lambda}  = q \sqrt{R-2\Lambda}$ for an appropriate metric tensor $g_{\mu\nu}$. This my be a hard problem, and it is the case with the corresponding  Schwarzschild-de Sitter metric in the nonlocal $\sqrt{dS}$ gravity. In the sequel, we will investigate the corresponding  Schwarzschild-de Sitter metric around static spherically symmetric massive body.

\section{The Schwarzschild-de Sitter Metric}

We are going to explore the Schwarzschild-de Sitter space-time metric in the case of  $\sqrt{dS}$ gravity given by its action \eqref{eq1.1}.

\subsection{General Consideration}

First, we want to consider some general aspects of the corresponding Schwarzschild-de Sitter space-time metric. To this end, we start from the usual expression for the Schwarzschild metric in the pseudo-Rimannian manifold
\begin{equation} \label{eq3.1}
  \mathrm ds^2 = - A(r)\mathrm dt^2 + B(r) \mathrm dr^2 + r^2\mathrm d\theta^2 + r^2\sin^2 \theta \mathrm d\varphi^2 ,     \qquad (c=1).
\end{equation}

Non-zero elements of metric tensor $g_{\mu\nu}(r)$ are:
\begin{align}\label{eq3.0}
 g_{00}(r) = - A(r), \quad g_{11}(r) = B(r), \quad  g_{22}(r) = r^2, \quad  g_{33}(r) = r^2 \sin^2{\theta}.
\end{align}
Non-zero components of the corresponding Christoffel symbol $\Gamma_{\mu\nu}^\alpha = \frac{1}{2} g^{\alpha\beta} \big(\partial_\mu g_{\nu\beta} + \partial_\nu g_{\nu\beta}  -  \partial_\beta g_{\mu\nu} \big)$ are:
\begin{align}
&\Gamma^0_{10} = \Gamma^0_{01}  = \frac{1}{2} \frac{A'}{A} , \qquad \Gamma^1_{00} = \frac{1}{2} \frac{A'}{B} ,  \qquad \Gamma^1_{11} = \frac{1}{2} \frac{B'}{B} , \qquad \Gamma^1_{22} = - \frac{r}{B} ,  \label{eq3.2a}   \\
&\Gamma^1_{33} = - \frac{r}{B} \sin^2{\theta} , \qquad \Gamma^2_{12} = \Gamma^2_{21} =  \frac{1}{r} ,  \qquad \Gamma^2_{33} = - \sin{\theta} \cos{\theta} ,   \label{eq3.2b} \\
&\Gamma^3_{13} = \Gamma^3_{31}= \frac{1}{r} , \qquad \Gamma^3_{23} = \Gamma^3_{32}= \frac{\cos{\theta}}{\sin{\theta}} ,
\end{align}
where ' denotes derivative with respect to radius $r.$
Non-zero components of the Ricci tensor $R_{\mu\nu} = R_{\mu\alpha\nu}^\alpha = \partial_\alpha \Gamma_{\mu\nu}^\alpha  - \partial_\nu \Gamma_{\mu\alpha}^\alpha   +  \Gamma_{\alpha\rho}^\alpha \Gamma_{\mu\nu}^\rho  - \Gamma_{\nu\rho}^\alpha \Gamma_{\mu\alpha}^\rho$  are:
\begin{align}
&R_{00} = \frac{A''}{2B} - \frac{A' B'}{4 B^2} - \frac{A'^2}{4A B} + \frac{A'}{r B}, \quad R_{11} = - \frac{A''}{2A} +\frac{A' B'}{4A B} + \frac{A'^2}{4 A^2} + \frac{B'}{r B} , \label{eq3.3a} \\
&R_{22} = - \frac{r A'}{2 AB} +\frac{rB'}{2B^2} - \frac{1}{B} + 1 , \quad R_{33} = \Big(- \frac{r A'}{2 AB} +\frac{rB'}{2B^2} - \frac{1}{B} + 1     \Big) \sin^2{\theta} ,  \label{eq3.3b}
\end{align}
where as a result of spherical symmetry $R_{33} = R_{22}  \sin^2{\theta} . $

The corresponding Ricci curvature $R = g^{\mu\nu} R_{\mu\nu}$ is
\begin{align} \label{eq3.4} R= \frac{2}{r^2}- \frac{1}{B(r)} \Big( \frac{2}{r^2}+\frac{2 A'(r)}{r A(r)}-\frac{A'(r)^2}{2 A(r)^2}
-\frac{2 B'(r)}{r B(r)}-\frac{A'(r)B'(r)}{2 A(r)B(r)}+\frac{A''(r)}{A(r)} \Big) .
\end{align}

Equation that should be solved is
\begin{align}
\Box u(r) = \frac 1{B(r)}\left(\triangle u(r) + \frac 12 \left(\frac {A'(r)}{A(r)} - \frac {B'(r)}{B(r)}\right)u'(r)\right) = q u(r), \quad u(r) = \sqrt{R-2\Lambda} \ ,  \label{eq3.5a}
\end{align}
where
\begin{align}  \label{eq3.5b} \triangle   = \frac{1}{r^2} \frac{\partial}{\partial r}\big[r^2  \frac{\partial  }{\partial r}  \big] = \frac{\partial^2}{\partial r^2} + \frac{2}{r} \frac{\partial}{\partial r}
\end{align}
is the Laplace operator in spherical coordinate $r$.

Note that in eigenvalue problem \eqref{eq3.5a} parameter $q$ has the same dimension as operator $\Box$, i.e. dimension is   $L^{-2}$. Since the cosmological constant $\Lambda$ also has dimension
$L^{-2}$, it is useful to write  $q = \zeta \Lambda$, where $\zeta$ is a dimensionless parameter.  Note also that nonlocal operator $\mathcal{F} (\Box)$, defined in  \eqref{eq1.2}, satisfies conditions $\mathcal{F} (q) = -1$ and $\mathcal{F}' (q) = 0$, introduced in \eqref{eq2.12}, if we take it as
\begin{align} \label{eq3.5c}
 \mathcal{F} (\Box) = e \Big[ a  \frac{\Box}{\zeta \Lambda}\ e^{\big(-\frac{\Box}{\zeta \Lambda}\big)} + b  \ \frac{\zeta \Lambda}{\Box}\ e^{\big(-\frac{\zeta \Lambda}{\Box}\big)}\Big] , \quad \text{where} \quad a + b = -1 .
 \end{align}

\subsection{Solutions}

Recall that in the local de Sitter case \eqref{eq2.4}, with  static spherically symmetric body of mass $M$, the Schwarzschild-de Sitter metric \eqref{eq3.1} is
\begin{align}
 A(r) = A_0(r) = 1 - \frac{\mu}{ r}  - \frac{\Lambda r^2}{3}, \quad B(r) = B_0(r) = \frac{1}{A_0(r)} = \frac{1}{1 - \frac{\mu}{ r}  - \frac{\Lambda r^2}{3}} , \qquad  \mu = \frac{2 G M}{c^2}  . \label{eq3.6}
\end{align}
It makes sense to suppose that solution of equation \eqref{eq3.5a} is of the form
\begin{align}
A(r) = A_0 (r) -  \alpha(r), \quad B(r) = \frac{1}{A_0(r) -  \beta(r)},  \label{eq3.7}
\end{align}
where $\alpha(r)$ and $\beta(r)$ are some dimensionless functions.
When $\zeta \to 0$ then nonlocal operator \eqref{eq3.5c} tends to zero and consequently  nonlocal de Sitter $\sqrt{dS}$ gravity model \eqref{eq1.1} becomes local. Hence, it must be that  $A(r) \to A_0 (r)$ and $B(r) \to B_0(r)$  when  $\zeta \to 0, $ that is  $ \alpha(r) \to 0$ and     $ \beta(r) \to 0$ as $\zeta \to 0 $ .

Replacing $A = A_0 - \alpha (r)$ and $B = \frac{1}{A_0 - \beta (r)}$ in scalar curvature R \eqref{eq3.4} and in operator $\Box$ of equation \eqref{eq3.5a}, we obtain
\begin{align}
  R &= \frac{2}{r^2} (1 -A_0 + \beta) + 2 \frac{A_0 - \beta}{A_0 -\alpha} (A'_0 - \alpha') \Big(\frac{1}{4}\frac{A'_0 - \alpha'}{A_0 - \alpha} - \frac{1}{r} \Big) \nonumber \\ &- 2 (A'_0 -\beta') \Big(\frac{1}{4}\frac{A'_0 - \alpha'}{A_0 - \alpha} + \frac{1}{r} \Big)  - \frac{A_0 -\beta}{A_0 - \alpha} (A''_0 - \alpha'') ,                         \label{eq3.8a} \\
  \Box u &= (A_0 -\beta ) \triangle u + \frac{1}{2} \Big[\frac{A_0 -\beta}{A_0 -\alpha} (A'_0 - \alpha') + A'_0 - \beta' \Big] u' = q u , \quad u = \sqrt{R - 2 \Lambda}  \label{eq3.8b}
\end{align}
If we  substitute expressions \eqref{eq3.8a} and \eqref{eq3.8b} in equation \eqref{eq3.5a} then we will get a differential equation with two unknown functions: $\alpha (r \sqrt{\zeta \Lambda})$  and $\beta (r \sqrt{\zeta \Lambda})$. It is obvious that we must have another equation which is a relation between functions $\alpha$ and $\beta$. Recall that in the local case holds $B_0(r) = \frac{1}{A_0 (r)}$. Hence, there is a sense to take also  $B(r) = \frac{1}{A (r)}$ in the nonlocal case and it yields
\begin{align}
R(r)  &= \frac{1}{r^2}\big[2 -2 A(r) - 4r A'(r)  - r^2 A'' (r)  \big] = \frac{1}{r^2} \frac{\partial^2}{\partial r^2}\big[r^2 \big(1-A(r) \big)\big] ,  \label{eq3.8c} \\
\Box u(r) &= A(r)\,u''(r)+ (A'(r) + \frac{2}{r}\;A(r))\,u'(r)= \frac{1}{r^2} \frac{\partial}{\partial r}\big[r^2 A(r) \frac{\partial u}{\partial r}  \big] . \label{eq3.8d}
\end{align}

 In fact, it means that we take
\begin{align}
   \beta (r) = \alpha (r)   .    \label{eq3.9}
\end{align}

Employing \eqref{eq3.9} in  \eqref{eq3.8a}  and  \eqref{eq3.8b}, we get
\begin{align}
R &= 4\Lambda + \frac{2 \alpha}{r^2} + \frac{4 \alpha'}{r} + \alpha'' ,     \label{eq3.10a} \\
\Box u &= (A_0 - \alpha) \triangle u +  (A'_0 - \alpha') u' = q u,  \qquad u = \sqrt{R - 2 \Lambda} .   \label{eq3.10b}
\end{align}

We are interested in finding function $\alpha (r)$ and the next step should be  substitution of
\begin{align} \label{eq3.11}
u = \sqrt{R - 2 \Lambda}  = \sqrt{2\Lambda + \frac{2 \alpha}{r^2} + \frac{4 \alpha'}{r} + \alpha'' }
\end{align}
into equation \eqref{eq3.10b}. In that case, one gets an ordinary nonlinear differential equation of the fourth order. Because of nonlinearity, it is a very difficult task to find the corresponding exact solution. After many attempts, we did not succeed  to find a reasonable exact solution and concluded  that a much more sophisticated approach is required. In the sequel of this paper we will turn our attention to the corresponding linear
differential equation.

\begin{figure}[H]
\begin{center}
\includegraphics[width=10.5 cm]{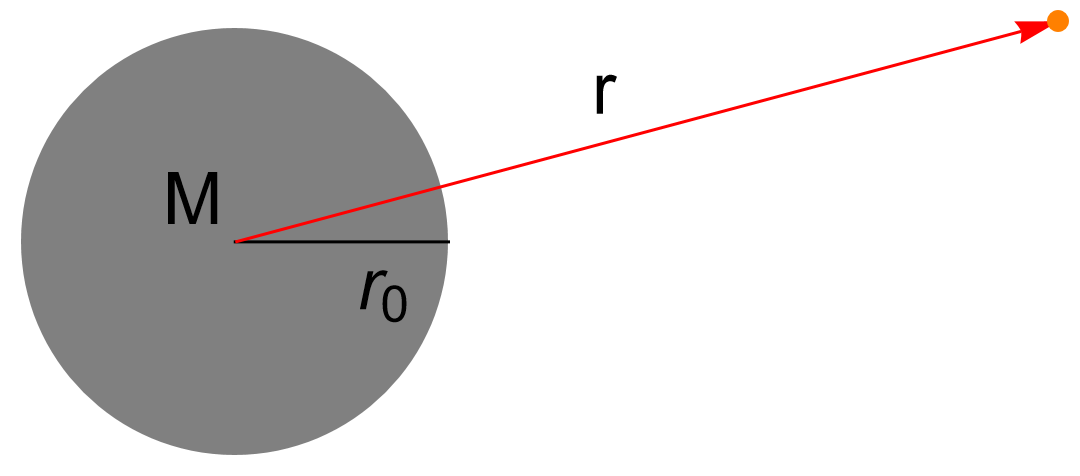}
\caption{We consider the Schwarzschild-de Sitter metric of nonlocal $\sqrt{dS}$ gravity  at the distances far from a spherically symmetric massive body. \label{fig1}}
\end{center}
\end{figure}

  It means we will limit ourselves  to studying the  Schwarzshild-de Sitter metric in weak gravity field approximation. Practically, it is like considering gravity field far from a massive body (see Figure~\ref{fig1}) so that the d'Alembertian $\Box$ can be replaced by the Laplacian $\triangle$ in equation
\eqref{eq3.10b}. In such case we will take  $A(r) \approx 1$ in \eqref{eq3.10b}, that is
\begin{align}
 A(r) = A_0 (r)- \alpha (r) = 1 - \frac{\mu}{ r}  - \frac{\Lambda r^2}{3} - \alpha(r) \approx 1, \label{eq3.12a}
\end{align}
what makes sense if the following is satisfied:
\begin{align}
  \frac{\mu}{r} \ll 1, \ \  \frac{\Lambda r^2}{3} \ll 1 , \ \  |\alpha(r)|  \ll 1 .   \label{eq3.12b}
\end{align}

Applying approximation \eqref{eq3.12a} in   \eqref{eq3.10b}, we get the following simple equation  linear in $u (r)$:
\begin{align}
\triangle u  = q u , \quad \text {that is} \quad    \frac{\partial^2 u}{\partial r^2} + \frac{2}{r} \frac{\partial u}{\partial r}
 = q u ,  \quad u = \sqrt{R- 2 \Lambda} . \label{eq3.13}
\end{align}

Note that requested function $\alpha (r)$ is contained in $u(r) = \sqrt{R- 2\Lambda}$, through \eqref{eq3.10a}: $R = 4\Lambda + \frac{2 \alpha}{r^2} + \frac{4 \alpha'}{r} + \alpha''$. Hence, the next step is linearization of $\sqrt{R- 2 \Lambda}$, and we get
\begin{align}
\sqrt{R- 2 \Lambda} = i \sqrt{2\Lambda - R}  \approx i \sqrt{2\Lambda} \Big( 1- \frac{R}{4 \Lambda} \Big) = - \frac{i}{2 \sqrt{2\Lambda}} (R -4\Lambda) \ , \quad \text{if} \quad \Big|\frac{R}{2\Lambda} \Big| \ll 1 .    \label{eq3.14}
\end{align}
It is worth noting that approximation  \eqref{eq3.14} was successfully applied in \cite{dimitrijevic10,dimitrijevic11}.

Now, replacing $\sqrt{R- 2 \Lambda}$ in \eqref{eq3.13} by $- \frac{i}{2 \sqrt{2\Lambda}} (R -4\Lambda)$, we get
\begin{align} \label{eq3.15}
 (R-4\Lambda)'' + \frac{2}{r} (R-4\Lambda)' = q (R-4\Lambda).
\end{align}
The next step is to replace scalar curvature $R$ by $R = 4\Lambda + \frac{2 \alpha}{r^2} + \frac{4 \alpha'}{r} + \alpha''$ (see \eqref{eq3.10a})
in \eqref{eq3.15} and we obtain
\begin{align} \label{eq3.16}
\alpha'''' + \frac{6}{r} \alpha''' + \frac{2}{r^2} \alpha'' - \frac{4}{r^3} \alpha' + \frac{4}{r^4} \alpha
= q (\alpha'' + \frac{4}{r} \alpha'+ \frac{2}{r^2} \alpha ) ,
\end{align}
that is a linear differential equation of the fourth order.

Equation \eqref{eq3.16} has a general solution of the form
\begin{align}\label{eq3.17}
  \alpha(r) = \frac{C_1}r + \frac{C_2}{r^2} + C_3 e^{-\sqrt q r}\left(\frac 1{q r} + \frac 2{q^{\frac 32} r^2}\right) +C_4 e^{\sqrt{q} r}\left(\frac1{qr}- \frac 2{q^{\frac 32}r^2}\right) , \quad q = \zeta \Lambda .
\end{align} There are four constants ($C_1, C_2, C_3, C_4$) and we have chosen them so that the appropriate particular solution $\alpha (r) \to 0$ when $\zeta \to 0$ and that it has some physical meaning.

To exclude term with $e^{\sqrt{q} r}$ in \eqref{eq3.17}, we took $C_4 =0$, since  this exponential function
increases indefinitely for very large values of $r$.

\subsubsection{Case\  $C_1 = -\frac{\delta}{\sqrt{q}},\ C_2 = \frac{2\delta}{q},\ C_3 = -\delta \sqrt{q}; \ \ C_4 = 0 $.}

In this case solution for $\alpha (r)$ is
\begin{align} \label{eq3.18}
\alpha (r) = -\frac{\delta}{\sqrt{q} r} \Big(1 + e^{-\sqrt{q} r}\Big)  + \frac{2\delta}{q r^2} \Big(1 - e^{-\sqrt{q} r}\Big) , \quad q=\zeta \Lambda ,
\end{align}
where $\delta$ is dimensionless parameter. Since integration constants $C_1, C_2, C_3$  are proportional to $\delta$ and $C_4 = 0$, by this way we reduced the number of parameters from 4 to 1. Altogether, we hove two free parameters ($\delta$ and $\zeta$) which should be determined from measurements.

 To see how $\alpha (r)$ behaves when $\zeta \to 0$, it is useful to expand exponential function  $e^{-\sqrt{q} r}$ into the Taylor series. We have
\begin{align}
\alpha (r) &= -\frac{\delta}{\sqrt{q} r} \Big(1 + e^{-\sqrt{q} r}\Big)  + \frac{2\delta}{q r^2} \Big(1 - e^{-\sqrt{q} r}\Big) \label{eq3.19a} \\
    &= \delta \Big(- \frac{2}{\sqrt{q} r} + 1 -  \frac{\sqrt{q} r}{2} + \frac{q r^2}{6}  - \frac{\sqrt{q} q r^3}{24} + \cdots    \Big)  \nonumber  \\
    &+ \delta \Big(\frac{2}{\sqrt{q} r} - 1 + \frac{\sqrt{q} r}{3} - \frac{q r^2}{12}  + \cdots    \Big) \label{eq3.19b}  \\
    &=  - \frac{\delta\sqrt{q} r}{6} +  \frac{\delta q r^2}{12} - \cdots =  - \frac{\delta\sqrt{\zeta \Lambda} r}{6} +  \frac{\delta \zeta \Lambda r^2}{12} - \cdots    \label{eq3.19c}
\end{align}
As it follows from \eqref{eq3.19c}, we  conclude that $\alpha (r) \to 0$ when $\zeta \to 0$.

According to \eqref{eq3.18}, we get
\begin{align} \label{eq3.20}
 A(r) = 1 - \frac{\mu}{r}  - \frac{\Lambda r^2}{3} +  \frac{\delta}{\sqrt{q} r} \Big(1 + e^{-\sqrt{q}\ r}\Big)  - \frac{2\delta}{q r^2} \Big(1 - e^{-\sqrt{q}\ r}\Big) ,  \quad  q = \zeta \Lambda ,
\end{align}
where $\mu = \frac{2 G M}{c^2} .$
When  $\zeta \to 0$,  obtained expression \eqref{eq3.20} for $A(r)$ tends to $A_0 (r)$, as necessary.

\subsection{The Rotation Curves of Spiral Galaxies}

The rotation curves of spiral galaxies have been the subject of intensive research motivated by the need to determine the amount and distribution of dark matter comparing to visible matter, see e.g. \cite{salucci,distribution,genzel,stability,stojkovic,capozziello} and references therein.

It is interesting to examine whether this $\sqrt{dS}$ gravitational model gives the possibility of describing the rotation curves of spiral galaxies. To this end, we should start with $A(r)$ given by \eqref{eq3.20} and present the corresponding gravitational potential $\Phi (r)$, which is
\begin{align}\label{eq3.21}
\Phi (r) &= \frac{c^2}{2} \big(1 - A(r)\big) = \frac{G M}{r} +\frac{\Lambda c^2 r^2}{6} + \frac{c^2}{2} \alpha (r)  \nonumber \\
      &= \frac{G M}{r} +\frac{\Lambda c^2 r^2}{6}   -\frac{\delta c^2}{2\sqrt{q} r} \Big(1 + e^{-\sqrt{q}\ r}\Big)  + \frac{\delta c^2}{q r^2} \Big(1 - e^{-\sqrt{q}\ r}\Big) .
\end{align} Note that here $\Phi (r)$ is the intensity of the gravitational potential.

The corresponding gravitational acceleration for potential \eqref{eq3.21} is
 \begin{align}\label{eq3.22}
 a_g(r)  = - \frac{\partial \Phi}{\partial r}  = \frac{G M}{r^2} -\frac{\Lambda c^2 r}{3} + \frac{\delta c^2}{\sqrt{q} r^2} \Big(\frac{2}{\sqrt{q} r} - \frac{1}{2}  \Big) - \frac{\delta c^2}{r}   \Big(\frac{1}{2} + \frac{3}{2 \sqrt{q} r}  + \frac{2}{q r^2}   \Big) e^{- \sqrt{q}\ r} .
\end{align}

Velocity of the rotation curve $\bar{v} (r)$ follows from equality  $\frac{\bar{v}^2 (r)}{r}  = a_g(r)$ and it is
\begin{align}\label{eq3.23}
 \bar{v} (r) = \sqrt{a_g(r)\ r} = c \sqrt{\frac{G M}{c^2 r} -\frac{\Lambda r^2}{3} + \frac{\delta }{\sqrt{q} r} \Big(\frac{2}{\sqrt{q} r} - \frac{1}{2}  \Big)  - \delta   \Big(\frac{1}{2} + \frac{3}{2 \sqrt{q} r}  + \frac{2}{q r^2}   \Big) e^{- \sqrt{q}\ r}}.
\end{align}
We checked the validity of the obtained formula for  the circular velocity \eqref{eq3.23} on two cases:  the Milky Way galaxy and the spiral galaxy M33. The values of parameters $\delta$ and $\zeta$ in\eqref{eq3.23} are estimated by best fitting of measured data using  the least-squares method. Since at large distances $r$, the velocity $\bar{v}$ weakly depends on mass variation, we employed only the mass of the black hole in the center of the galaxy. Namely, the central mass can be taken up to $M \sim 10^8 M_\odot$ and there will be no significant changes in the calculated circular velocity $\bar{v} (r)$ at very large $r$.

\subsubsection{Milky Way case.}

\begin{figure}[H]
\begin{center}
\includegraphics[width=10.5 cm]{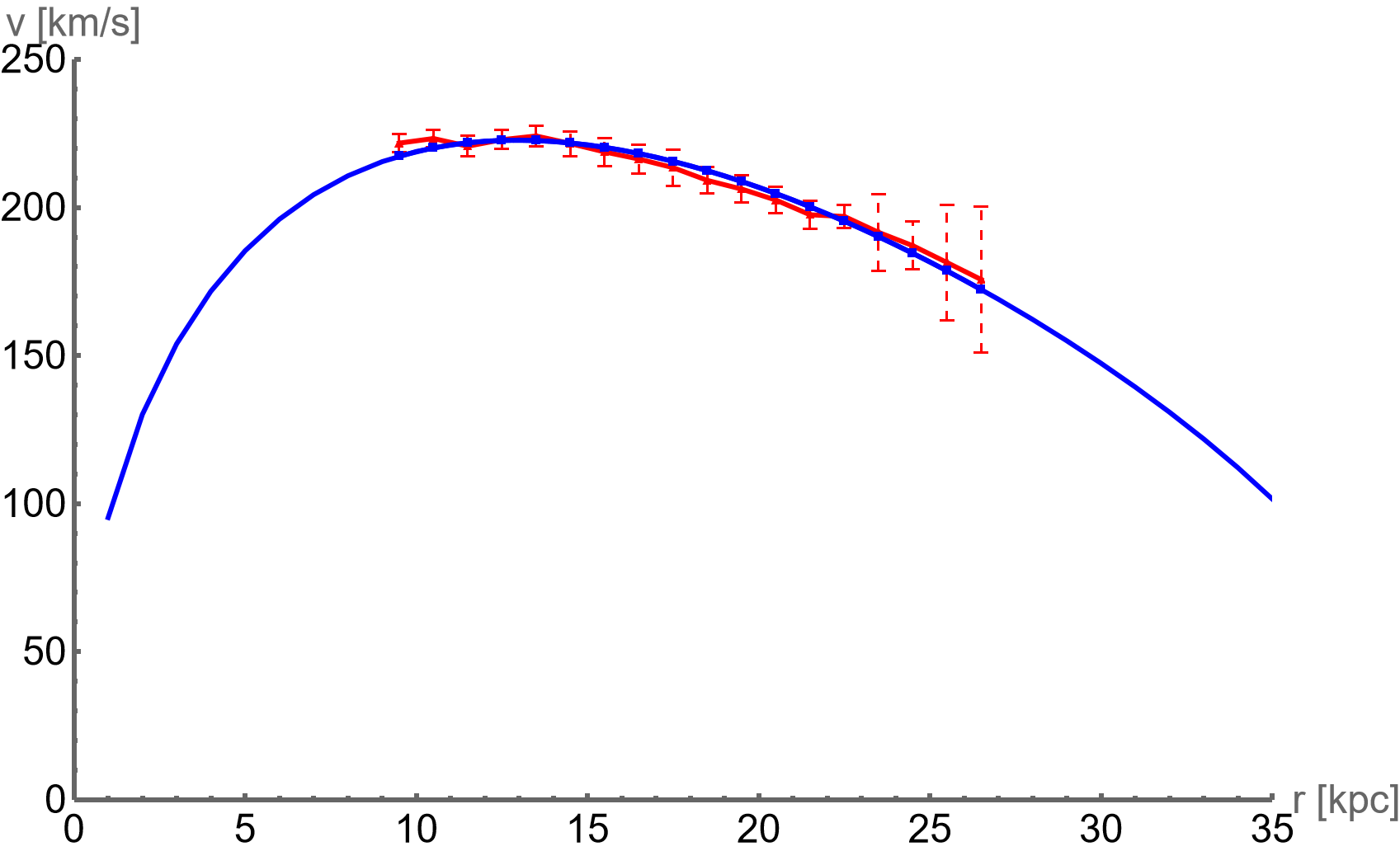}
\caption{Rotation curve  for the Milky Way galaxy. Red points are measured observational values \cite{milkyway} and blue line is computed $\bar{v}(r)$ by formula \eqref{eq3.23}, where $\delta = 1.9\times 10^{-5}$, $\zeta = 4.4\times 10^{10}$, $\Lambda = 10^{-52} \text{m}^{-2}$ and $M =4.28 \times 10^6 M_\odot$.\label{fig2}}
\end{center}
\end{figure}

The Milky Way rotation curve data have taken from recent paper \cite{milkyway}, where the Keplerian decline in the rotation curve is detected.
Measured data for distance $r$, velocity $v$ and velocity error $\Delta v$ are obtained by {\it Gaia} telescope and they are presented in Table 1, see \cite{milkyway}. In this table also are presented computed velocity $ \bar{v}$ using \eqref{eq3.23} and relative error $\delta v = \frac{|v - \bar{v}|}{v}$. A pictorial comparison of measured and calculated velocities is presented in Figure~\ref{fig2}.

\begin{table}[H]
\caption{Milky Way rotation curve data, from \cite{milkyway} and this work. \label{tab1}}
\begin{center}
\begin{tabularx}{0.6\textwidth}{CCCCC}
\hline
			$r\,$ [kpc]	& $v\,$ [km/s]	& $\Delta v\,$ [km/s]  & $\bar v\,$ [km/s] &\textbf{relative error $\delta v$ [\%]} \\
\hline
 9.5  & 221.75 & 3.17 & 217.36 & 1.98 \\
 10.5 & 223.32 & 3.02 & 220.19 & 1.40 \\
 11.5 & 220.72 & 3.47 & 221.93 & 0.55 \\
 12.5 & 222.92 & 3.19 & 222.72 & 0.09 \\
 13.5 & 224.16 & 3.48 & 222.66 & 0.67 \\
 14.5 & 221.60 & 4.20 & 221.85 & 0.11 \\
 15.5 & 218.79 & 4.75 & 220.37 & 0.72 \\
 16.5 & 216.38 & 4.96 & 218.28 & 0.88 \\
 17.5 & 213.48 & 6.13 & 215.63 & 1.01 \\
 18.5 & 209.17 & 4.42 & 212.47 & 1.58 \\
 19.5 & 206.25 & 4.63 & 208.83 & 1.25 \\
 20.5 & 202.54 & 4.40 & 204.77 & 1.10 \\
 21.5 & 197.56 & 4.62 & 200.29 & 1.38 \\
 22.5 & 197.00 & 3.81 & 195.42 & 0.80 \\
 23.5 & 191.62 &12.95 & 190.17 & 0.75 \\
 24.5 & 187.12 & 8.06 & 184.57 & 1.36 \\
 25.5 & 181.44 &19.58 & 178.62 & 1.55 \\
 26.5 & 175.68 &24.68 & 172.32 & 1.91 \\
\hline
		\end{tabularx}
\end{center}
\end{table}

\subsubsection{Spiral galaxy M33 case.}

\begin{figure}[H]
\begin{center}
\includegraphics[width=10.0 cm]{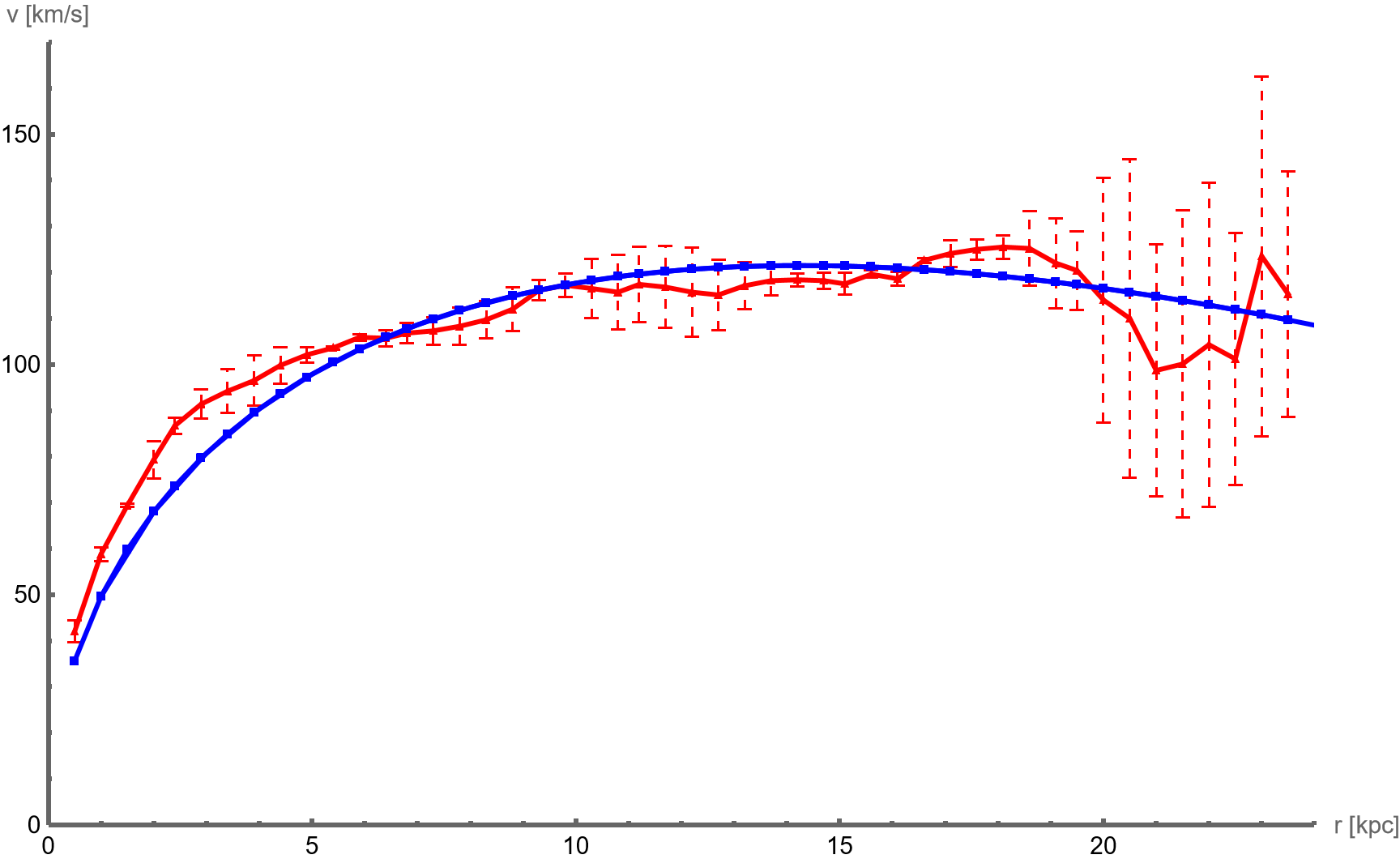}
\caption{Rotation curve  for  spiral galaxy M33. Red points are measured observational values and blue line is computed $\bar{v}(r)$ by formula \eqref{eq3.23}, where $\delta = 5.7\times 10^{-6}$, $\zeta = 3.62\times 10^{10}$,  $\Lambda = 10^{-52} \text{m}^{-2}$  and $M =1.5 \times 10^3 M_\odot$ .\label{fig3}}
\end{center}
\end{figure}

We have used data for the galaxy Messier 33, based on observations obtained at the Dominion
Radio Astrophysical Observatory and  presented in \cite{M33}. We have compared measured rotation velocity $v (r)$ \cite{M33} with the computed $\bar{v} (r)$ using formula \eqref{eq3.23}. Measured and computed data are presented in Table \ref{tab2} and illustrated in Figure \ref{fig3}.

\begin{table}[H]
\begin{center}
\caption{$M33$ galaxy data, from \cite{M33} and this work.\label{tab2}}
\begin{tabularx}{\textwidth}{CCCCC|CCCCC}
\hline
			$r\,$ [kpc]	& $v\,$ [km/s]	& $\Delta v\,$ [km/s]  & $\bar v\,$ [km/s] &\textbf{relative error $\delta v$ [\%]} & $r\,$ [kpc]	& $v\,$ [km/s]	& $\Delta v\,$ [km/s]  & $\bar v\,$ [km/s] &\textbf{relative error $\delta v$ [\%]}\\
\hline
 0.5 & 42.0 & 2.4 & 35.62 & 15.18 & 12.2 & 115.7 & 9.6 & 120.69 & 4.31 \\
 1.0 & 58.8 & 1.5 & 49.61 & 15.63 & 12.7 & 115.1 & 7.7 & 121.05 & 5.17 \\
 1.5 & 69.4 & 0.4 & 59.83 & 13.79 & 13.2 & 117.1 & 5.1 & 121.30 & 3.58 \\
 2.0 & 79.3 & 4.0 & 68.02 & 14.22 & 13.7 & 118.2 & 3.2 & 121.45 & 2.75\\
 2.4 & 86.7 & 1.8 & 73.59 & 15.12 & 14.2 & 118.4 & 1.4 & 121.50 & 2.62\\
 2.9 & 91.4 & 3.1 & 79.64 & 12.86 & 14.7 & 118.2 & 1.8 & 121.47 & 2.76\\
 3.4 & 94.2 & 4.8 & 84.90 &  9.88 & 15.1 & 117.5 & 2.4 & 121.38 & 3.30\\
 3.9 & 96.5 & 5.5 & 89.51 &  7.25 & 15.6 & 119.6 & 0.8 & 121.19 & 1.33\\
 4.4 & 99.8 & 3.9 & 93.58 &  6.23 & 16.1 & 118.6 & 1.5 & 120.93 & 1.96\\
 4.9 & 102.1 & 1.7& 97.21 & 4.80  & 16.6 & 122.6 & 0.5 & 120.59 & 1.64\\
 5.4 & 103.6 & 0.4 & 100.44& 3.05& 17.1 & 124.1 & 2.9 & 120.17 & 3.16\\
 5.9 & 105.9 & 0.7 & 103.32& 2.44& 17.6 & 125.0 & 2.2  & 119.69 & 4.24\\
 6.4 & 105.7 & 1.7 & 105.90& 0.19& 18.1 & 125.5 & 2.5 & 119.15 & 5.06\\
 6.8 & 106.8 & 2.2 & 107.76& 0.90& 18.6 & 125.2 & 8.1 & 118.54 & 5.32\\
 7.3 & 107.3 & 3.0 & 109.86 & 2.39& 19.1 & 122.0 & 9.8 & 117.87 & 3.38\\
 7.8 & 108.3 & 4.0 & 111.73 & 3.17& 19.5 & 120.4 & 8.5 & 117.29 & 2.58\\
 8.3 & 109.7 & 4.0 & 113.34 & 3.37& 20.0 & 114.0 & 26.6 & 116.52 & 2.21\\
 8.8 & 112.0 & 4.8  & 114.86 & 2.55& 20.5 & 110.0 & 34.6 & 115.70 & 5.18\\
 9.3 & 116.1 & 2.2 & 116.15& 0.04& 21.0 & 98.7 & 27.4 & 114.82 & 16.33 \\
 9.8 & 117.2 & 2.5 & 117.27& 0.06& 21.5 & 100.1 & 33.4 & 113.89 & 13.77 \\
 10.3 & 116.5 & 6.5 & 118.24 & 1.49& 22.0 & 104.3 & 35.2 & 112.91 & 8.25\\
 10.8 & 115.7 & 8.1 & 119.07 & 2.91& 22.5 & 101.2 & 27.4 & 111.88 & 10.56\\
 11.2 & 117.4 & 8.2 & 119.63 & 1.90& 23.0 & 123.5 & 39.1 & 110.81 & 10.27\\
 11.7 & 116.8 & 8.9 & 120.22 & 2.93& 23.5 & 115.3 & 26.7 & 109.69 & 4.86\\
\hline
		\end{tabularx}\end{center}
\end{table}

\section{Discussion and Concluding Remarks}

This paper presents the results of our research regarding  the Schwarzschild-de Sitter metric of the nonlocal $\sqrt{dS}$ gravity model \eqref{eq1.1}. We found the Schwarzschild-de Sitter metric in the form of $A(r)$ \eqref{eq3.20},  what corresponds to  the weak gravity approximation and the linearization of nonlinear differential equation \eqref{eq3.10b}. The obtained results were tested on the rotation curves of the Milky Way and the spiral galaxy M33.
The calculated and measured values of circular velocities are in  good agreement.

Some additional explanations should be given to some parts of these investigations. First, we need to clarify why the weak gravitational field approximation works well here. On the one hand, we derived the Schwarzschild-de Sitter metric  away from the massive spherically symmetric body.
And on the other hand, we applied the obtained formula for the circular motion of the test body to the circular velocities in spiral galaxies far from their centers where the black hole is located. Recall that the rotation curves were observed in the domain: 9.5 --26.5 kpc for the Milky Way galaxy \cite{milkyway} and 0.5 --23.5 kpc for the M33 galaxy \cite{M33}. In the Lambda Cold Dark Matter model, it is assumed that dark matter plays an important role in the mentioned domains. However, there is no dark matter in our nonlocal model. The good agreement between observational measurements and theoretical predictions tells us that the role of dark matter can be played by the nonlocality in the presence  of the cosmological constant $\Lambda$ in the $\sqrt{dS}$ gravity model.

Regarding the applicability of the obtained formula for the circular velocity \eqref{eq3.23} at smaller distances, such as the solar system, the following should be noted.   The circular velocity $\bar{v} (r)$   depends on three terms:  (i) $\frac{G M}{c^2 r}$, (ii)  $-\frac{\Lambda r^2}{3}$ and (iii)  $ \frac{\delta }{\sqrt{q} r} \Big(\frac{2}{\sqrt{q} r} - \frac{1}{2}  \Big)  - \delta   \Big(\frac{1}{2} + \frac{3}{2 \sqrt{q} r}  + \frac{2}{q r^2}   \Big) e^{- \sqrt{q}\ r}$. The third term depends linearly on $\delta$ and with a fixed $q (= \zeta \Lambda)$ its value can be controlled by choosing the appropriate value of $\delta$.  One can always take a small enough value of $\delta$, e.g. $\delta < |8.2 \times 10^{-14}|$, so that the first term has a dominant role, since the second term has an important meaning only at distances of the size of the visible universe.  Therefore, the velocity formula \eqref{eq3.23} is also valid for the solar system.

The main new results presented in this article can be summarized as follows.
\begin{itemize}
\item	In the approximation of the weak gravitational field, a  fourth-order linear differential equation for the Schwarzschild-de Sitter metric  was obtained \eqref{eq3.16}.
\item	A general solution  \eqref{eq3.17} of  equation \eqref{eq3.16} was found.
\item	A particular solution of $\alpha (r)$ was found \eqref{eq3.18} such that it satisfies the necessary condition that it tends to zero when the nonlocality vanishes.
\item  The obtained theoretical formula for circular velocity \eqref{eq3.23} was tested on the rotation curves of two spiral galaxies: the Milky Way and M33.
 The agreement between the calculated and measured circular velocities is good, especially for the Milky Way, see Figures 2 and 3 and the corresponding tables. To our knowledge, this is the first good description of ``the Keplerian decline in the Milky Way rotation curve'' by some modified gravity model.
\end{itemize}

In summary, it can be said that the presented results in this paper are encouraging and deserve further research, especially taking into account the mass distribution in spiral galaxies using \cite{fitting}. Bearing in mind also
 previously obtained results on the evolution of the universe \cite{PLB,JHEP}, where the effects that are usually attributed to dark energy and dark matter can be described by the nonlocality of the gravity model $\sqrt{dS}$, we will continue with the further study of this model of nonlocal de Sitter gravity.


This research was partially funded by  the Ministry of Science, Technological Development and Innovation of the Republic of Serbia, grant numbers: 451-03-66/2024-03/ 200104 and 451-03-65/2024-03/ 200138.  This paper is also based on work partially related to the COST Action CA21136, ``Addressing observational tensions in cosmology with systematics and fundamental physics'' (CosmoVerse) supported by COST (European Cooperation in Science and Technology).

The authors are grateful to D. Stojkovic, P. Jovanovic and D. Borka for helpful discussions.



\end{document}